\documentclass[12pt]{amsart}
\usepackage{soul}
\input{amssymb.sty}

\makeatletter
\newcommand\ackname{Acknowledgements}
\if@titlepage
  \newenvironment{acknowledgements}{%
      \titlepage
      \null\vfil
      \@beginparpenalty\@lowpenalty
      \begin{center}%
        \bfseries \ackname
        \@endparpenalty\@M
      \end{center}}%
     {\par\vfil\null\endtitlepage}
\else
  \newenvironment{acknowledgements}{%
      \if@twocolumn
        \section*{\abstractname}%
      \else
        \small
        \begin{center}%
          {\bfseries \ackname\vspace{-.5em}\vspace{\z@}}%
        \end{center}%
        \quotation
      \fi}
      {\if@twocolumn\else\endquotation\fi}
\fi
\makeatother

\begin{document}

\title[Non-commutative Digital Signatures]
{Non-commutative Digital Signatures}
\author[D. Kahrobaei]{Delaram Kahrobaei}
\address{CUNY Graduate Center and NYCCT, City University of New York}%
\email{DKahrobaei@GC.Cuny.edu}
\thanks{Research of the first author was partially supported by the Office of Naval Research grant N000141210758, PSC-CUNY grant from the CUNY research foundation, as well as the City Tech foundation.}
\author[C. Koupparis]{Charalambos Koupparis}
\address{CUNY Graduate Center, City University of New York}%
\email{ckoupparis@GC.Cuny.edu}
\begin{abstract}
The objective of this work is to survey several digital signatures proposed in the last decade using non-commuta\-tive groups and rings and propose a digital signature using non-commutative groups and analyze its security.
\end{abstract}

\maketitle

\section{Introduction to Digital Signatures}
We start by describing digital signatures using an analogy of a signed message (document) from the non-digital world, whereby a person signs a document, seals it in an envelope and mails it to a recipient. Upon receipt of the envelope the recipient opens and examines the document, specifically the signature, to verify the authenticity of the document and that the author was in fact the expected sender of the envelope. 

Similarly a digital  signature scheme provides a way for each user to sign messages so that the signatures can later be verified by anyone else. To be precise, each user creates a matched pair of private and public signatures for the message (using the signer's public key). The verifiers can convince themselves that the message contents have not been altered since the message was signed. Furthermore, the signer cannot later deny having signed the message, since no one but the signer possesses his private key. 
The recipient can perform the inverse operations of opening the letter and verifying the signature. Such signature schemes for electronic mail are already quite widespread today (see \cite{GB08}). This is often cited as one of the most fundamental and useful inventions of modern cryptography. 

\section{The Ingredients of Digital Signatures}
We follow the notation of Goldwasser and Bellare in their MIT lecture notes (for further reading and definitions see \cite{GB08}). 
A digital signature scheme within the public key framework, is defined as a tuple of algorithms $(G, \sigma, V)$. 
The key generation algorithm $G$ takes as input a security parameter $\alpha$ and outputs $P$ and $S$, a public key and a secret key respectively. The signing algorithm $\sigma$ takes as input a security parameter $\alpha$, the secret key $S$ and a
 message $m$. The output produced is a string $s$, the signature of the message $m$. Finally, the verification algorithm $V$ 
 when given the public key $P$, a digital signature $s$, and a message $m$, returns either true or false indicating whether or not
 the signature is valid.

\subsection{Classical digital signatures}
We briefly mention a couple of classical digital signatures, again following \cite{GB08}.
\subsubsection{RSA Digital Signature Scheme}
The RSA digital dignature scheme is based on the RSA cryptosystem. The public key consists of a pair of integers $(n, e)$ where $n$ is the product of two large primes, $e$ is relatively prime to $\phi(n)$ ( $\phi$ is Euler's totient function). The secret key, $d$, is chosen  such that $ed = 1$ mod $\phi(n)$. One signs a message by computing the signature $\sigma(m) =m^d$ mod $n$. To verify that this is a valid signature one raises the signature to the power $e$ and compares it to the original message.

\subsubsection{El Gamal Digital Signature Scheme}
The El Gamal digital signature scheme is based on the Diffie-Hellman key exchange (DHKE) problem, and the difficulty of solving this problem.  Presently, it is suggested that the best approach to tackling the DHKE problem is to first solve the discrete log problem. However, it is unknown whether computing a discrete log is as hard as solving the Diffie-Hellman problem. The DHKE problem upon input a prime $p$, a generator $g$ of the group $\mathbb{Z}^*_p$ and the two elements $g^x$ 
and $g^y$ (for $x,y \in \mathbb{Z}$), seeks to determine $g^{xy}\mod p $.

%Here is the idea of the scheme:
%\begin{itemize}
%	\item Public key: A triple $(y, p, q)$, where $y=g^x$ mod $p$, where $p$ is  prime and $g$ is a generator 
%	of $\mathbb{Z}^*_p$.\\
%	\item Secret key: $x \in \mathbb{Z}$ such that $y=g^x$ mod $p$.\\
%	\item Signing: The signature of message $m$ is a pair $(r, s)$ such that $0 \not= r, s \not= p-1$ and $g^m = y^r r^s$ mod $p$.\\
%	\item Verifying: Check that $g^m = y^r r^s$ mod $p$ actually holds.	
%\end{itemize}
%
%In order to generate a pair $(r, s)$ which constitutes a valid signature, the signer begins by choosing a random number $k$ such that $0 \not=k\not=p-1$ and $GCD(k, p-1)= 1$. Define $r=g^k$ mod $p$. Next we need to determine an $s$ such that $g^m = y^r r^s = g^{xr+ks}  \mod p$. Looking at the exponents in this equation we have the relationship $m = xr + ks$ (mod $p-1$). Solving this for $s$ yields $s = (m - xr)k^{-1}$ mod $p-1$. The signature of $m$ is then the pair $(r, s)$. It is obvious that if an attacker could solve the discrete logarithm problem, he could break the scheme completely by computing the secret key $x$ (or $y$) from the public information. Moreover, care needs to be taken in choosing the pseudo random number generator, since if an attacker finds $k$ for one message, he can solve the discrete logarithm problem.
%
\subsubsection{Schnorr Digital Signature Scheme}
The Schnorr signature algorithm's security is based on the intractability of certain discrete logarithm problems \cite{Schnorr91}. 
This signature scheme is considered one of the simplest digital signature schemes to be provably secure in a random oracle model. It is both efficient and allows for the generation of short signatures.

%Here is the idea of the scheme: Select a prime $q$, select $1 \leq a \leq q-1$; and compute $y=g^a$ %mod $p$.
%\begin{itemize}
%	\item Public key: (p, q, g, y)$ \\
%	\item Secret key: .\\
%	\item Signing: .\\
%	\item Verifying: .	
%\end{itemize}

\subsection{Non-commutative digital signatures using non-commutative groups and rings}
\subsubsection{Braid Groups}
In 2002 Ko, Choi, Cho and Lee \cite{ko-new} proposed a digital signature using braid groups where they assume the 
conjugacy search problem is hard, but the conjugacy decision problem is feasible. 

In 2009 Wang and Hu \cite{WH09} proposed a new digital signature based on a non-commutative group. Their signature scheme is based on the root extraction problem over braid groups.

We note that in general the conjugacy search problem in braid group based schemes are susceptible to length-based attacks (see \cite{MU07}, \cite{GKTTV06} and \cite{HT02}) and as such may not be suitable as platforms for non-commutative digital signatures. 

\subsubsection{Division Semirings}

Another example of generating digital signatures over non-commutative algebraic objects was given by Anjaneyulu, Reddy 
and Reddy in 2008 \cite{ARR08}. They consider polynomials over non-commutative division semirings. They assume that the 
computational Diffie-Hellman problem is hard in their setup. Additionally their signature also relies on the difficulty of the generalized symmetrical decomposition (GSD) problem as applied to their rings. 

%Let $R$ be a non-commutative division semiring, define $P_a=\{f(a) | f(x)$ $\in \mathbb{Z}_{>0}[x]\}$, where $a \in R$. Then the 
%GSD problem applied here is called the Polynomial Symmetrical Decomposition (PSD) problem. I.e., given $(a,x,y) \in R^3$ 
%and $m, n \in \mathbb{Z}$, find $z \in P_z$ such that $y=z^mxz^n$.
%
%The public information is the tuple $(S,m,n,M,H)$, where S is the non-commutative division semiring, $m,n \in \mathbb{Z}$ 
%and $H$ is a cryptographic hash function which maps the message space $M$ to $S$. 
%
%Alice publishes $(p,q,y)$ where $p,q$ are chosen randomly from $S$ and y is generated by choosing a random polynomial 
%$f(x) \in \mathbb{Z}_{>0}[x]$ such that $0 \ne f(p) \in S$, and computing $y=f(p)^mqf(p)^n$.
%
%To generate a signature Alice choses another random $h(x) \in \mathbb{Z}_{>0}[x]$ such that $h(p)\in S$ and she defines 
%\[u= h(p)^mqh(p)^n.\] She then computes
%\begin{align*}
%r &= f(p)^mH(M)uf(p)^n \\
%s &= h(p)^mrh(p)^n \\
%\alpha &= h(p)^mrf(p)^n \\
%\beta &= f(p)^mH(M)h(p)^n \\
%v_1 &= h(p)^mH(M)h(p)^n
%\end{align*}
%She then publishes $(u, s, \alpha, \beta, v_1)$ as the signature of the message $M$. 
%
%Bob verifies this is a valid signature
%by computing
%\[v_2=\alpha y^{-1}\beta.\]
%He accepts the signature if and only if
%\[u^{-1}v_1=s^{-1}v_2.\]
%

The authors propose that their signature is both secure against data forging of the message and against existential forgery. However, we believe that both these claims may be incorrect. In their scheme if someone replaces the valid message 
$M$ with a forged message $M_f$, then the signature already sent would be valid. Although $M$ is used in creating the 
signature, it is not needed in verification of the signature. Hence the verification test will succeed. 

For existential forgery one is required to produce a valid 
signature for any message of their choosing. As such, one can at will choose parameters that satisfy their verification algorithm. 

%$\alpha, y, \beta$ which forces $v_2=\alpha y^{-1}\beta$. We must then satisfy 
%$u^{-1}v_1=s^{-1}v_2$, hence choose any 2 of the 3 remaining unknowns and solve the equation for the thrid, without loss of generality assume we chose 
%$u$ and $s$, then $v_1 = us^{-1}v_2$. Hence a signature that will be accepted for any $M$ may be genrated using this process. 

\subsubsection{General Non-Commutative Rings}

In order to limit the ability of a third party to verify the validity of a signature, Chaum and van Antwerpen ~\cite{CA90} introduced
 the notion of \textit{undeniable signatures}. Like a digital signature, undeniable signatures depend on the signer's public key as well as on the message signed. However, verification can only  be achieved by interacting with the legitimate signer through a \textit{confirmation protocol}. This method also allows the signer to deny the signature. In particular, if the signer refuses to deny, 
 or fails to deny the signature, then the signature is assumed to be legitimate. Furthermore, as the signer's cooperation is 
 required for verification of the signature they are protected from verification attempts by unauthorized third parties.
 
% The setup requires a non-commutative group, $G$, in which the Conjugacy Search Problem (CSP) and the Conjugacy Decision Problem (CDP) are hard. Additionally we require two subgroups $A$ and $B$ of $G$,  such that $[A,B]=1$. Alice chooses random $g \in G$ and $a \in A$ and computes $x=aga^{-1}$ The public key is then $(g,x)$. To sign a message $m$, the message is hashed as $y=H(m)$ and Alice computes $S=aya^{-1}$.
% 
%The confirmation protocol is then executed. Upon receipt of a signature pair $(m,\hat{S})$ the verifier Bob chooses $b\in B$ 
%and sends the challenge $Q=b\hat{S}xb^{-1}$ to Alice. Alice chooses $c,d \in G$ and sends the response
% $R=dgc(a^{-1}Qa)c^{-1}d^{-1}$ to Bob.
% 
% Bob then sends $b$ to Alice and she verifies that $Q=b\hat{S}xb^{-1}$. Alice then sends $c, d$ to Bob and he verifies that
% $R=dgcb(yg)b^{-1}c^{-1}d^{-1}$. If this verification holds, then Bob accepts $\hat{S}$ as a valid signature from Alice.

\section{A non-commutative digital signature}
Let $G$ be an infinite finitely presented group with exponential growth rate, such that there is no known polynomial-time algorithm for solving the conjugacy search problem. In the signature we use $f$ represents a simple mapping function $f:G\rightarrow \{0,1\}^*$, which maps our group to some binary representation that can be digitally encoded. We will also be using a collision-free hash function $H$ which maps into $G$. We note that for our algorithm Alice's public key will have to be updated/changed periodically depending on the 
number of messages she transmits. 

\begin{itemize}
\item{\bf Setup} The signer, Alice, chooses a group element $g$, a private key $s \in G$ and an integer $n \in \mathbb{N}$. We note that in our scheme $n$
should be chosen to be a highly composite number, $n=\prod_{k=1}^l p_k^{e_k}$, where $p_k$ are prime and $e_k \in \mathbb{N}$.
She then computes 
$x=g^{ns}$ and publishes $x$. Note, when exponentiating with an element of $h\in G$ we are representing conjugation, $g^h=h^{-1}gh$. Furthermore, the centralizer of $g$ should be trivial, i.e. the set of group elements commuting with $g$ should consist of only the identity.
\item{\bf Key generation:} The signer wishes to sign the message $m$ which is a bit string. She picks $t$ uniformly at random from $G$, a random 
factorization of $n = n_i n_j$, and computes the key $y=g^{n_it}$.
\item{\bf Signature:} To generate the signature $\sigma$ compute the following:
\begin{align*}
h & = H(m || f(y)) \\
\alpha & = t^{-1}shy 
\end{align*}
Alice then publishes her signature $\sigma = (y, \alpha,n_j)$ and the message $m$.
\item{\bf Verification:} To verify the signature compute  $h' = H(m || f(y)) $. The signature is valid and accepted if and only if
\begin{align*}
y^{n_j\alpha} &= x^{h'y} 
\end{align*}
\end{itemize}

\subsection{Security Analysis of the Signature Protocol} 

We note that the idea for this algorithm was generated by Schnorr's digital signature for commutative groups. In particular, 
the use of string concatenation and a hash function were borrowed from this scheme.

\subsubsection{Completeness}
Given a signature generated by Alice $(y, \alpha,n_j)$, and the public key $x$, Bob will always accept the signature 
as valid following the verification algorithm.

 First Bob computes $h' = H(m || f(y)) = h$. He then verifies the equation $y^{n_j\alpha} = x^{h'y}$. The left hand side yields 
 \[ y^{n_j\alpha} = y^{n_jt^{-1}shy}=g^{n_in_jtt^{-1}shy}=g^{nshy}=x^{hy}.\] As $h'=h$ the equation is valid, hence the protocol is complete.
 
\subsubsection{Data forging}
 Suppose that the forger Eve replaces the valid message to be signed, $m$, with a forged message, $m_f$. Then when Bob computes $h' = H(m_f || f(y) ) \ne H(m ||f(y)) = h$ he won't be able to verify that $y^{n_j\alpha} = x^{h'y} \ne x^{hy}$. This equation in 
 general doesn't hold unless there is a collision in the hash function for the particular choices of $(m, y, f)$, 
 which is unlikely given our assumptions about $H$. 
 
\subsubsection{Existential Forgery}
Suppose Eve wishes to sign a forged message $m_f$. She would then have to generate a valid signature 
$\sigma=(y_f,\alpha_f,n_{j_f})$ which passes the verification algorithm. It is here where is becomes necessary to use an exponent $n$. For if $n=1$ then the verification reads $y^{\alpha} = x^{h'y} \Rightarrow y^{\beta h'y}=x^{h'y} \Rightarrow y^\beta=x$. Hence choosing $\beta$ determines $y$, which in turn gives us $h$ and hence $\alpha$. Combining all this yields an existentially forged signature 
for  any $m$.

Repeating the above in our case yields $y^{n_j\beta} = x$. In order to solve this equation $y, \beta$ and $n_j$ must be determined. 
A priori it is not clear how this may be done. One may proceed by choosing 2 of the 3 unknowns and solving for the third. 
In this case, if $\beta$ is the last parameter left, then we are left to solve the CSP problem, which is we know to be difficult 
for a given platform group. Hence one must 
choose $\beta$. If we next choose $n_j$, then we need to solve $y^{n_j}=x^{\beta^{-1}}$. We are not guaranteed a solution 
of this equation in general as this implies the existence of and the ability to compute roots in the underlying group. Hence this forces us to choose both $\beta$ and $y$, which again means we need to solve a DH problem which may or may not have a 
solution and is already computationally hard.

Based on the above we believe that existential forgery of this protocol is not possible unless one already knows a root of $x$. It 
turns out that this can be done once Alice has sent out a message and its signature. One can determine an $n_j^{th}$ 
root of $x$ by  computing $y^{\alpha y^{-1}h^{-1}}$. In order to stop Eve from forging a message using this $n_j$ 
Alice needs to keep a public list updated with the $n_j$'s she has used to far. 
If we receive a message with an $n_j$ already used then we know it must not be from Alice.

Another option is an adaptive chosen ciphertext attack, where Eve gets to submit messages of her choosing for signing. 
Again this method of attack is unlikely to succeed as the most Eve will obtain is a distribution of $t^{-1}s$, which is random and 
should yield no information about $s$ nor $t$. In addition, Eve will be receiving information about $n$ as well, however, 
as suggested before, once Alice has exhausted a small list of factorizations we recommend switching to a new $x$ and $n$. This 
switch can be prolonged if careful choices are made in the factorization of $n$. In particular we can specifically choose not to 
include certain primes in the integer $n_j$ that is published, or even to restrict the exponents of the primes used in $n_j$. 

 \subsubsection{Soundness}
One method of breaking the security requires the eavesdropper to recover $s$, $t$ or $n$. 
Since $g$ nor $n$ was never published, nor needed, there isn't a clear method of starting a CSP attack. 
You would either have to be able to attack the random algorithm which generates $t$ and hence obtain information 
about $t^{-1}s$, or there would have to be some method of attacking the hash 
function. Hence the security of this signature generation protocol relies on the appropriate choice of hash function and the 
method by which one obtains random group elements. Care must also be taken as to how the elements are transmitted. Since an 
eavesdropper can always read back $s^{-1}t$, for random $t$, we must make sure that this doesn't leak any information about $s$.

\subsection{Proposed Platforms}
We advocate using platform groups for which the conjugacy search problem is hard. Such non-commutative groups have been discussed in \cite{MSU11}. In particular, any group which has been deemed secure against length based attacks and other attacks may be used. Such groups include polycyclic groups as they have been 
proposed for cryptography in \cite{EK04}, \cite{AK09} and \cite{KK06}. Garber, Kahrobaei and Lam in \cite{GKL} have done some experiments which shows that well-chosen polycyclic groups with high Hirsch length are secure against length based attacks.
For a survey on non-commutative group-based cryptography see \cite{FHKR11} and \cite{MSU11}.

\begin{acknowledgements}
The authors would like to thank the anonymous referees for their helpful and insightful comments. 
\end{acknowledgements}

\bibliographystyle{amsplain}
\bibliography{XBib}
\end{document}